\documentclass[amsmath,superscriptaddress,twocolumn,prl,showpacs]{revtex4}

\usepackage{amssymb}
\usepackage{graphicx,psfrag}
\usepackage{natbib}
\usepackage{multirow}
\usepackage{amsmath}

\begin{document}

\title{Electron spin dynamics in strongly correlated metals}

\author{Bal\'{a}zs D\'{o}ra}
\email{dora@pks.mpg.de}
\affiliation{Budapest University of Technology and Economics, Institute
of Physics and Condensed Matter Research Group of the Hungarian Academy of
Sciences, H-1521 Budapest, Hungary}
\affiliation{Max-Planck-Institut f\"ur Physik
Komplexer Systeme, N\"othnitzer Str. 38, 01187 Dresden,
Germany}
\author{Ferenc Simon}
\email{ferenc.simon@univie.ac.at}
\affiliation{Budapest University of Technology and Economics, Institute
of Physics and Condensed Matter Research Group of the Hungarian Academy of
Sciences, H-1521 Budapest, Hungary}

\begin{abstract}

The temperature dependence of the electron spin life-time, $T_1$ and the $g$-factor are anomalous in alkali 
fullerides (K,Rb)$_3$C$_{60}$, which cannot be explained by the canonical Elliott-Yafet theory. These materials 
are archetypes of strongly correlated and narrow band metals. We introduce the concept of "complex electron 
spin resonance frequency shift" to treat these measurables in a unified manner within the Kubo formalism. 
The theory is
applicable for metals with nearly degenerate conduction bands and large momentum scattering even with an
anomalous temperature dependence and sizeable residual value.

\end{abstract}


\maketitle


Spintronics, i.e. the use of the spin degree of freedom of electrons for information processing \cite{FabianRMP}, is a rapidly 
developing field. Its research is motivated by the orders of magnitude longer conservation of the electron 
spin 
alignment in metals as compared to their momentum conservation time. The survival of spin orientation, characterized by $T_1$, determines the time window for spin manipulation. The $g$-factor determines the
magnetic energy of the electrons and characterizes the conditions for magnetic resonance manipulations of spins.

The seminal theory of Elliott and Yafet (EY) was developed in the 1950's to explain the electron spin relaxation and $g$-factor in metals \cite{Elliott,YafetReview}. These parameters are measured with conduction electron spin resonance (CESR) \cite{KipKittel1952}. In CESR, the metal is placed in a magnetic field and is irradiated with microwaves. Resonant microwave absorption occurs when the irradiation energy equals the Zeeman splitting of the electron spins.

The EY theory is based on the presence of spin-orbit coupling as it mixes a near lying band with the conduction band states. Thus 
conduction band states are admixtures of spin up and down causing i) a change in the magnetic energy of conduction electrons (i.e. a 
$g$-factor shift, $\Delta g$) and ii) allowed transitions between the two spin states 
(i.e. spin relaxation) when the electron is 
scattered on defects or phonons with momentum relaxation time $\tau$. Elliott and Yafet showed with first order time-dependent 
perturbation theory that:

\begin{eqnarray}
\Gamma_{\text{spin}}&=&\alpha_1\left(\frac{L}{\Delta}\right)^2\Gamma \label{ElliottRelaxation}, \\ 
\Delta g=g-g_0&=&\alpha_2\frac{L}{\Delta} \label{Elliottgfactor}, 
\end{eqnarray}

\noindent where $\alpha_{1,2}$ are band structure dependent constants of the order of unity, $g_0=2.0023$ is the $g$-factor of the free electron, $L$ is the matrix element of the SO coupling between the near
lying and the conduction band state, separated by an energy gap $\Delta$, $\Gamma_{\text{spin}}=\hbar/T_1$, $\Gamma=\hbar/\tau$.

Eqs. \ref{ElliottRelaxation} and \ref{Elliottgfactor} are summarized in the Elliott-relation: $\Gamma_{\text{spin}}=\alpha \Delta g^2 \Gamma$, where the constant $\alpha \approx 1..10$. Typically $\Delta g^2 \approx 10^{-4}..10^{-7}$ thus $T_1$ is orders of magnitude longer than $\tau$. The experimentally accessible measurables are the CESR line-width: $\Delta B=\Gamma_{\text{spin}}/\hbar \gamma $ (where $\gamma/2 \pi= 28.0$ GHz/T is the electron gyromagnetic factor) and the resistivity, $\rho \propto \Gamma$. Thus the Elliott-relation establishes that the CESR line-width and the resistivity are proportional, which enabled an experimental verification for most elemental metals by Monod and Beuneu \cite{BeuneuMonodPRB1978,BeuneuMonodPRB1979}.

\begin{figure}
{\includegraphics[]{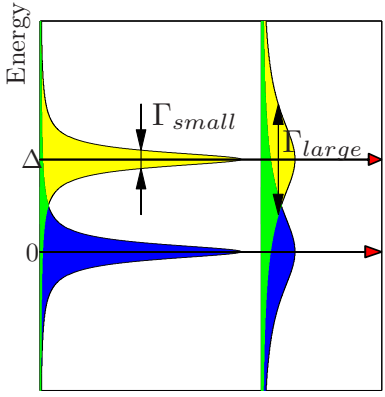}}
\caption{The schematic view of the single particle spectrum at small $\Gamma( \ll \Delta$) addressed by the EY theory and at 
large $\Gamma( \sim \Delta$), which calls for the generalized EY theory. Due 
to the large overlap (green area), the effective band-gap is reduced.}
\label{spectrum}
\end{figure}

Much as the Elliott-Yafet theory has been confirmed, it is violated in MgB$_2$ at high temperatures as therein $\Delta B$ and $\rho$ 
are not proportional \cite{SimonPRL2001}, which was explained \cite{EYgeneralizedPRL2008} by extending the EY theory for the case of 
rapid momentum scattering: $\Delta \approx \Gamma$. The EY theory neglects the role of $\Gamma$ with respect to $\Delta$ as in usual 
metals $\Gamma \approx 1-10$ meV and $\Delta \approx 1-10$ eV. The need for a generalized EY theory is even more demanding in alkali 
fullerides 
whose conduction band is composed of a triply degenerate molecular orbital making a small effective $\Delta$ 
and where large electron-phonon coupling and structural disorder gives rise to a large $\Gamma$. The situation is sketched in Fig.
\ref{spectrum}.

\begin{figure*}
\includegraphics[width=1\hsize]{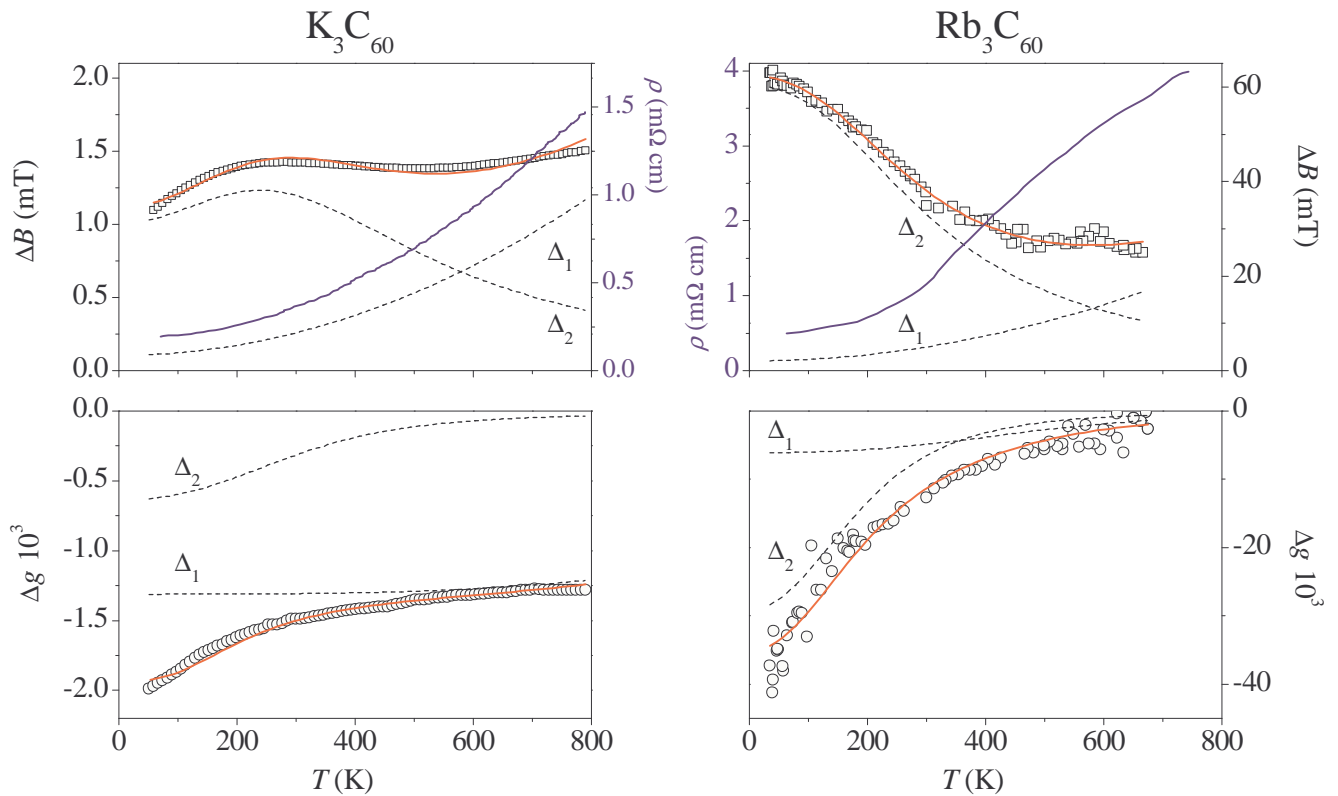}
\caption{(Color online) The temperature dependent resistivity (solid blue curve from Ref. \cite{HouSSC1995}),
CESR line-width ($\Box$), and $g$-factor shift ($\bigcirc$) (Refs. (\cite{JanossyPRL1993,PetitPRB1996,NemesPRB})
in the fullerides. ESR data on Rb$_3$C$_{60}$ above 300 K is from the present work. Solid curves are calculated
with the model explained in the text. Dashed curves are the contributions from the  different $\Delta_1$ and $\Delta_2$.
Note the different scales for the two compounds.}
\label{Fig2_fits}
\end{figure*}

The $\Delta B$ and the $g$-factor are shown in Fig. \ref{Fig2_fits}. for K$_3$C$_{60}$ and Rb$_3$C$_{60}$ and show that the 
$g$-factor is temperature dependent and the CESR line-width does not follow the resistivity in violation of the EY theory. Here, we 
introduce the concept of the "complex ESR frequency shift" which allows a simultaneous treatment of $T_1$ and the $g$-factor within 
the Kubo formalism. We also include a significant residual momentum scattering rate (dirty limit) and we find that the theory 
explains quantitatively the experimental observables, which enables to establish 
the  "generalized Elliott-relation".

We prepared Rb$_3$C$_{60}$ powder samples by a conventional solid state reaction method \cite{DresselhausFullerenes} using stoichiometric amounts of sublimation purified
C$_{60}$ and elemental Rb to study its ESR properties up to high temperatures which has not been performed yet. ESR data is available for K$_3$C$_{60}$ for the 4-800 K temperature range \cite{NemesPRB}. A sharp superconducting transition and high Meissner shielding in a 1 mT magnetic field (measured with microwave conductivity) together with the observation of the characteristic CESR signal of Rb$_3$C$_{60}$ \cite{JanossyPRL1993} attest the high quality of the material.
A sample of 10 mg sealed under helium in a quartz tube was measured in a commercial X-band (9 GHz) ESR spectrometer
in the 100-700 K temperature range.


The description of $T_1$ and $g$-factor is based on a two-band model Hamiltonian, $H=H_0+H_{\text{SO}}$, where:

\begin{equation}
\begin{split}
H_0=\sum_{k,\nu,s}\left[\epsilon_\nu(k)+\hbar \gamma B
s\right]c^+_{k,\nu,s}c_{k,\nu,s}+H_{\text{scatt}},\\
H_{\text{SO}}=\sum_{k,\nu\neq\nu',s,s'} L_{s,s'}(k)c^+_{k,\nu,s}c_{k,\nu',s'}
\label{ModelHamiltonian}
\end{split}
\end{equation}

\noindent Here $\nu, \nu'=1 \text{ or }2$ are the band, $s,s'$ are spin indices, $L_{s,s'}$ is the SO coupling, and $B$ is the
magnetic field along the $z$ direction. $H_{\text{scatt}}$ is responsible for the finite $\tau$. We use the Mori-Kawasaki
formula \cite{mori,oshikawa} to determine the $T_1$ and  $\Delta g=\hbar\Delta\omega_{\text{L}}/\mu_{\text{B}} B$ which allows
to introduce the "complex ESR frequency shift":

\begin{gather}
\Delta \widetilde{\omega}_{\text{ESR}}:=\Delta\omega_{\text{L}}-\frac{i}{T_1}=\dfrac{-\langle 
[P,S^-]\rangle+G^R_{PP^+}(\omega_{\text{L}})}
{2\langle S_z\rangle},
\label{deltagMK}
\end{gather}

\noindent where $\langle S_z\rangle$ is the expectation value of the spin along the magnetic field, $\omega_{\text{L}}=\gamma 
B$ 
is
the Larmor frequency,
and $G^R_{PP^+}(\omega)$ is the retarded Green's function of the $P$ and $P^+$ pair with $\hbar P=[H_{\text{SO}},S^+]$.
Eq. \ref{deltagMK} is evaluated with the unperturbed Hamiltonian, $H_0$, to yield the lowest non-vanishing correction
due to SO coupling as it is much smaller than the temperature or the band-gap. We note that Eq. \ref{deltagMK} is analogous to
the complex conductivity with para- and diamagnetic terms \cite{mahan}.

To enable an analytic calculation \cite{mahan}, we assume two linear bands with different Fermi velocities, both crossing 
the Fermi
energy
at different points and separated by $\Delta$, which yields:

\begin{gather}
\Gamma_{\text{spin}}(=\hbar \gamma \Delta B)=\left\langle \frac{L^2}{\Delta^2+\Gamma^2}\Gamma\right\rangle_{\text{FS}},
\label{ESRT1}
\end{gather}

\begin{gather}
\Delta g=\frac{1}{\pi k_{\text{B}}T}\left\langle 2 L_z \textmd{Im}\Psi^{\prime}\left(\frac 12+\frac{\Gamma+\text{i}\Delta}{2\pi
k_{\text{B}}T}\right)\right\rangle_{\text{FS}}
\label{ESRgfactor}
\end{gather}

\noindent where the $\langle\dots\rangle_{\text{FS}}$ means Fermi surface (FS) averaging, $\Psi'(x)$ is the first derivative of Euler's
digamma function,
and all parameters $L$, $\Delta$, and $\Gamma$ are taken on the FS.
$L_z=L_{\uparrow,\uparrow}-L_{\downarrow,\downarrow}$ and $L^2=L_z^2 +2|L_{\downarrow,\uparrow}|^2$. Eq.
\ref{ESRgfactor}.
can be simplified if $2 \pi k_{\text{B}}T \lesssim \Gamma$ to give

\begin{gather}
\Delta g= \left\langle \frac{2 L_z \Delta}{\Delta^2+\Gamma^2}\right\rangle_{\text{FS}}
\label{ESRgfactor_simplified}
\end{gather}

Eqs. \ref{ESRT1} and \ref{ESRgfactor}. return the corresponding EY results (Eqs. \ref{ElliottRelaxation}-\ref{Elliottgfactor}) when
$\Gamma\ll \Delta$ and is regarded as a generalization of the EY theory. If Eqs. \ref{ESRT1}. and \ref{ESRgfactor_simplified}. 
can be handled with isotropic band-band separation and SO coupling, the generalized Elliott-relation is:

\begin{gather}
\Gamma_{\text{spin}}= \alpha \Delta g^2 \Gamma \text{ }\left(1+ \frac{\Gamma^2}{\Delta^2}\right)
\end{gather}
 \noindent which returns the conventional formula when $\Gamma \ll \Delta$.

\begin{table}[t!]
\caption{Residual, $\rho_0$, and high temperature, $\rho(T_{\text{h}})$ resistivity and the corresponding momentum scattering rates, $\Gamma$, for K$_3$C$_{60}$ ($T_{\text{h}}=790$ K) and Rb$_3$C$_{60}$ ($T_{\text{h}}=700$ K) from Ref. \cite{HouSSC1995}. Plasma frequencies are from Ref. \cite{GunnRMP1997}. The coefficient $A$, defined in the text, is also given.}
\begin{tabular*}{0.5\textwidth}{@{\extracolsep{\fill}}cccccccccccc}
\hline \hline
& $\omega_{\text{pl}}$&$\rho_0$ & $\Gamma_0$& $\rho(T_{\text{h}})$&$\Gamma(T_{\text{h}})$&$A$\\
& (eV)& (m$\Omega$cm)& (meV)&(m$\Omega$cm)&(meV)& (meV/K$^2$)\\
\hline
K$_3$C$_{60}$ & 1.2&0.2 &  39 & 1.5 & 285& 3.94$\cdot 10^{-4}$ \\
Rb$_3$C$_{60}$ & 1.1&0.5 &  81 & 4.0 & 650&13.1$\cdot 10^{-4}$\\
\hline \hline
\label{Resistivity_Parameters_Table}
\end{tabular*}
\end{table}

We proceed to analyze the line-width and $g$-factor in alkali doped fullerides. 
Knowledge of the temperature dependent $\Gamma$ is required, which we determine from 
 resistivity data on single crystals by Hou \textit{et al.} \cite{HouSSC1995} (solid 
blue curves in Fig. \ref{Fig2_fits}.) using theoretical plasma frequencies, $\omega_{\text{pl}}$ 
through $\rho=1/\epsilon_0 \omega_{\text{pl}}^2 \tau$ (where $\epsilon_0$ is the electric constant). 
These compounds are unique in two aspects: i) the resistivity is high and it follows a quadratic 
temperature dependence up to the highest available temperatures, and ii) the residual resistivity 
is also high and it is not related to a residual impurity concentration but is intrinsic. The high 
value and quadratic temperature dependence of $\rho$ was explained by the coupling of electrons to 
the high energy intramolecular phonons, whereas the large residual value was associated with an 
inherent disorder of the C$_{60}$ ball orientation (the so-called merohedral disorder) \cite{GunnRMP1997,GunnNat}. 
The latter is related to the frustrated nature of the C$_{60}$ icosahedra with respect to the cubic molecular crystal lattice. 
The parameters of \endnote{Although there is an apparent saturation of $\rho$ in Rb$_3$C$_{60}$ above 500 K, the quadratic fit lies within 15 \% of the experimental value.} $\Gamma(T)=\Gamma_0+A \cdot T^2$, are given in Table \ref{Resistivity_Parameters_Table}.

 The $g$-factor is independent of the temperature and $\Delta B$ increases monotonously in the EY theory. In contrast, both 
measurables deviate from the expected behavior in K$_3$C$_{60}$ and Rb$_3$C$_{60}$ as shown in Fig. \ref{Fig2_fits}: $|\Delta g|$ 
decreases with increasing temperature and $\Delta B$ does not follow the resistivity. Most surprisingly, $\Delta B$ \emph{decreases} 
on increasing temperature in Rb$_3$C$_{60}$. The generalized EY theory shows that $\Delta\approx \Gamma$ explains the saturating and 
decreasing $\Delta B$ and decreasing $|\Delta g|$ (see Fig. \ref{spectrum}). However, a small $\Delta$ alone can not explain the data 
and Fermi surface parts, 
where $\Delta> \Gamma$, are also present.

\begin{table*}[t!]
\caption{Best fitting parameters used to simulate the experimental line-width and $g$-factor data in K$_3$C$_{60}$ and
Rb$_3$C$_{60}$. Note the different relative DOS in the two materials.}
\begin{ruledtabular}
\begin{tabular*}{0.95\textwidth}{@{\extracolsep{\fill}}ccccccc} 
&$|L|$ (meV)& $L_z$ (meV)&\multicolumn{2}{c}{$\Delta$ (eV)}   &  \multicolumn{2}{c}{$N$ (\%)}  \\
&  & &1&2& 1&  2 \\
\hline
K$_3$C$_{60}$ & 0.67(1) & -0.63(1)&0.94(3) & 0.047(2) & 3.0(2) & 97(2) \\
Rb$_3$C$_{60}$ &1.10(2)&-3.7(1)&0.35(3)& 0.050(2)& 31(2)& 69(2)\\
\end{tabular*}
\end{ruledtabular}
\label{Linewidth_FitParameters_Table}
\end{table*}

To handle the complicated Fermi surface of the fullerides the simplest possible way, we assume that the FS consists of two parts: one 
with large and another with small $\Delta$ ($\Delta_1$ and $\Delta_2$, respectively) with different relative density of states 
(DOS), $N_1$ and $N_2$. We assume uniform $L$ and $\Gamma$. This allows to approximate the FS averages in Eqs. \ref{ESRT1}. and
 \ref{ESRgfactor}. with a sum of two components.  In Fig. \ref{Fig2_fits}. we show the calculated $\Delta B$ and $\Delta g$ with 
this two component sum. The fit parameters are given in Table \ref{Linewidth_FitParameters_Table}. Calculations with a single 
$\Delta$ fail to account for the data in both compounds. We judge that the fits are in reasonable agreement with the experiment, 
given the simplifications to the general theory in terms of a two component sum. We note that Adrian \cite{AdrianPRB1996} suggested 
that a relation similar to Eq. \ref{ESRT1} explains $\Delta B$ for Rb$_3$C$_{60}$, however apart from a qualitative hint, no attempt 
for a quantitative analysis was made.

The obtained $\Delta$ values are compatible with the known small, $< 1 \text{ eV}$ band-width of alkali fullerides \cite{GunnRMP1997}. 
The conduction band is derived from the triply degenerate $t_{\text{1u}}$ molecular orbital of C$_{60}$, whose degeneracy is lifted in 
the fulleride crystal. However, the merohedral disorder prevents the knowledge of the band structure and we infer band structure 
properties from our analysis. 
It shows that on some parts of the FS, the band crossing it has a neighboring band as close as 50~meV. On other parts, the nearest
neighboring band is as close as 0.35-1~eV.

The two compounds only differ in the relative amount of such FS parts: for K$_3$C$_{60}$ parts with 
small $\Delta$ dominate whereas for Rb$_3$C$_{60}$ the relative DOS for the two types of Fermi surfaces are almost equal.

Only the $L/\Delta$ ratio is available in the EY theory but the correlated metals allow measurement of $\Delta$ and $L$ independently. We note that both $L$ and $L_z$ contain band structure dependent constants of the order of unity. The negative sign of $L_z$ reflects the electron-like (as opposed to hole-like) character of the conduction states, which is a common situation in e.g. alkali metals. $|L|$ and $L_z$ are unequal already in the EY framework \cite{BeuneuMonodPRB1978,BeuneuMonodPRB1979}, which is also the situation herein. For both compounds the SO couplings are about three orders of magnitude smaller than the corresponding values for elemental K (0.26 eV) and Rb (0.9 eV) (Ref. \cite{YafetReview}). This is due to the weak character of the alkali orbitals in the conduction band of C$_{60}$ \cite{ErwinPedersonPRL1991,AndreoniPRB1995}. On average, the corresponding SO coupling parameters are $\sim 3.7$ times larger in Rb$_3$C$_{60}$ than in K$_3$C$_{60}$ which is in good agreement with the $\sim 3.5$ ratio found for the elemental metals.

We comment on the application related aspects of the extended EY theory. In the $\Gamma \gg \Delta$ limit, we observe that
$\Gamma_{\text{spin}}\approx L^2/\Gamma$, which is formally identical to the result of the so-called Dyakonov-Perel mechanism
\cite{DyakonovPerel}. The latter occurs for semiconductors without inversion symmetry (i.e. large Dresselhaus SO coupling) 
and relatively long $\tau$, i.e. when the electron spins precess around the internal SO magnetic fields between momentum scattering events. The
spin-diffusion length, $\delta_{\text{spin}}= v_{\text{F}}\sqrt{\tau T_1}/3$ (where $v_\text{F}$ is the Fermi velocity) tends to a
constant in the above limit and $\delta_{\text{spin}}= v_{\text{F}}\hbar/3L$. $\delta_{\text{spin}}$ is one of the most 
important parameters for spintronics as it describes the geometrical path for spin transport \cite{FabianRMP}.

Line-width, resistivity, and $v_{\text{F}}=1.8\cdot 10^{5}$
m/s \cite{GunnRMP1997} data for K$_3$C$_{60}$ gives a relatively long $\delta_{\text{spin}}\approx$ 180 nm at 800 K, 
which is competitive at this high temperature to noble metals such as Cu (100 nm), Ag (180 nm), or
Au (40 nm) even if the Fermi velocities are an order of magnitude longer for the latter compounds. This demonstrates that molecular 
metals with nearly degenerate metallic bands are potentially interesting for spintronics applications.

Finally, we summarize in what sense the novel description points beyond the EY theory. The EY theory was developed and tested in metals where i) band-band separation is much larger than the quasi-particle scattering rate energy, i.e $\Delta \gg \Gamma$, ii) the residual $\Gamma$ is essentially zero, iii) $\Gamma(T)$ is linear with the temperature. Alkali fullerides do not possess any of these properties, still our theory accounts for the measured CESR parameters. The current description is applicable to a broad range of metals thus we expect that it will lead to smart design of materials for future spintronics devices.

FS acknowledges the Bolyai programme of the Hungarian Academy of
Sciences for support. Work
supported by the Hungarian State Grants (OTKA) No. F61733,  K72613, and NK60984.


\bibstyle{apsrev}
\bibliography{ESR_T1}

\end{document}